\documentstyle[12pt,
epsfig]{article}
\relax

\def\be{\begin{equation}}
\def\ee{\end{equation}}
\def\bs{\begin{subequations}}
\def\es{\end{subequations}}
\def\calm{{\cal M}}
\def\calmb{{\bar{\cal M}}}

\def\calmb{{\bar{\cal M}}}
\def\sb{{\bar{\sigma}}}

\def\lx{\lambda}
\def\ex{\epsilon}

\newcommand{\sx}{\sigma}

\newcommand{\Ab}{{\bar A}}
\newcommand{\Omb}{{\bar \Omega}}
\newcommand{\Hb}{{\bar H}}
\newcommand{\rb}{{\bar r}}

\newcommand{\lb}{{\bar \lx}}

\newcommand{\tb}{{\bar t}}
\newcommand{\ttb}{{\bar t}}
\newcommand{\fb}{{\bar f}}

\newcommand{\Rb}{{\bar R}}

\newcommand{\rhb}{\bar{\rho}}

\newcommand{\kb}{\bar{k}}

\def\be{\begin{equation}}
\def\ee{\end{equation}}
\def\bs{\begin{subequations}}
\def\es{\end{subequations}}
\newcommand{\een}{\end{subequations}}
\newcommand{\ben}{\begin{subequations}}
\newcommand{\beq}{\begin{eqalignno}}
\newcommand{\eeq}{\end{eqalignno}}

\newcommand\fverb{\setbox\pippobox=\hbox\bgroup\verb}
\newcommand\fverbdo{\egroup\medskip\noindent%
                        \fbox{\unhbox\pippobox}\ }
\newcommand\fverbit{\egroup\item[\fbox{\unhbox\pippobox}]}
\newbox\pippobox

\newcommand{\bea}{\begin{eqnarray}}
\newcommand{\bdm}{\begin{displaymath}}
\newcommand{\edm}{\end{displaymath}}

\newcommand{\eea}{\end{eqnarray}}
\def\sc{\bar{H}_i }
\newcommand{\bt}{\bar{t}}
\newcommand{\bR}{\bar{R}}
\newcommand{\brr}{\bar{r}}
\newcommand{\sbA}{\sqrt{\bar{A}}}
\newcommand{\sbAz}{\sqrt{\bar{A}^{(0)}}}
\newcommand{\sbAf}{\sqrt{\bar{A}^{(1)}}}
\newcommand{\sbAs}{\sqrt{\bar{A}^{(2)}}}

\newcommand{\br}{\bar{r}}

\begin{document}

\begin{center}
{ \Large \bf
Analytical Estimate \\ of the Effect of Spherical Inhomogeneities \\
on Luminosity Distance
and Redshift} 
\\
\vspace{1.5cm}
{\Large 
N. Brouzakis  and N. Tetradis 
} 
\\
\vspace{0.5cm}
{\it
Department of Physics, University of Athens,\\
University Campus, Zographou 157 84, Greece
} 
\end{center}
\vspace{3cm}
\abstract{
We provide an analytical estimate of the effect of a spherical inhomogeneity on light beams
that travel through it. We model the interior of the inhomogeneity through the 
Lemaitre-Tolman-Bondi metric. 
We assume that the beam source is located outside the inhomogeneity. 
We study the relative deviations of travelling time, redshift, 
beam area and luminosity distance from their values in a homogeneous cosmology. 
They depend on the ratio $\Hb=H r_0$ of the 
radius $r_0$ of the inhomogeneity to the horizon distance $1/H$. 
For an observer located at the center, the deviations are of order $\Hb^2$.
For an observer outside the inhomogeneity, the deviations of crossing time and
redshift are of order $\Hb^3$. The deviations of beam area and luminosity distance
are of order $\Hb^2$.
However, when averaged over all possible locations of the 
observer outside the inhomogeneity, they also become of order $\Hb^3$. We discuss
the implications for the possibility of attributing the observed cosmological
acceleration to the emergence of large-scale structure.
}
%\maketitle

\newpage

{\bf Introduction}:
The cause of the perceived acceleration of the present cosmological expansion 
has not been identified yet. An interesting possibility, that does not require
the introduction of new ingredients to Standard Cosmology, is that the growth of
inhomogeneities in the matter distribution affects the astrophysical observations
similarly to accelerated expansion in a homogeneous background. In particular, the
luminosity distance of faraway sources may be increased because of the propagation of
light through inhomogeneous regions before reaching the observer.

An unambiguous way to examine this possibility is through the study of the 
transmission of light in an exact inhomogeneous background. The analytical modelling of 
the Universe can only be approximate, and depends on the scale of the assumed
inhomogeneities. At length scales above
${\cal O} (10)\, h^{-1}$ Mpc the density contrast is at most of ${\cal O}(1)$. 
A popular choice for the background is based on the Lemaitre-Tolman-Bondi (LTB) 
metric \cite{ltb}. The background has spherical symmetry, but can 
be inhomogeneous along the radial direction. The metric can be matched to the 
Friedmann-Robertson-Walker (FRW) metric at a certain radius $r_0$.  
There are two possible choices for the location of the observer, which are consistent
with the isotropy of the Cosmic Microwave Background. 
a) He/she could be
located in the interior of the inhomogeneity, near its center \cite{central}. 
b) He/she could be located in the homogeneous region, with the light travelling
across several inhomogeneities during its propagation from source to observer
\cite{biswas2,brouzakis,brouzakis2}. 

In both cases, the size of the inhomogeneity $r_0$ determines its effect 
on quantities such as redshift and source luminosity distance.
The relevant quantity is the dimensionless ratio $\Hb=r_0 H$ of $r_0$ to
the horizon distance $1/H$. Consistency with observations requires that $\Hb$
be of ${\cal O}(10^{-2})$, even though values larger by an order of magnitude
have also been advocated for the explanation of the supernova data \cite{central}.
In the following we use perturbation theory in $\Hb$ in 
order to determine the dependence of the photon
redshift and source luminosity distance on $\Hb$, for both possible
locations of the observer. 
\\

{\bf Gravitational background}:
The LTB metric can be written in the form
\begin{equation}
ds^{2}=-dt^2+\frac{R'^2(t,r)}{1+f(r)}\,dr^2+R^2(t,r)d\Omega^2,
\label{metrictb}
\end{equation}
where $d\Omega^2$ is the metric of a two-sphere,
the prime denotes differentiation with respect to $r$, and
$f(r)$ is an arbitrary function.
The function $R(t,r)$ describes the location of a shell of matter marked by $r$
at the time $t$. Through an appropriate rescaling it can be chosen to satisfy
$R(0,r)=r$.

The Einstein equations reduce to
\begin{eqnarray}
\dot{R}^2(t,r)&=&\frac{1}{8\pi M^2}\frac{\calm (r)}{R}+f(r)
\label{tb1} \\
\calm'(r)&=&4\pi R^2 \rho(t,r) \, R',
\label{tb2} \end{eqnarray}
where the dot denotes differentiation with respect to $t$, and
$G=\left( 16 \pi M^2 \right)^{-1}$.
The generalized mass function $\calm(r)$ of the pressureless fluid 
with energy density $\rho(t,r)$ can be chosen arbitrarily. 

We parametrize the energy density at some arbitrary initial time $t_i=0$
as $\rho_i(r)=\rho(0,r)=\left( 1+\ex(r)\right)\rho_{0,i}$.
The initial energy density of the homogeneous background is
$\rho_{0,i}$. If the size of the inhomogeneity is
$r_0$, the matching with the homogeneous metric in the exterior requires
$4\pi\int_0^{r_0} r^2 \ex(r) dr=0$,
so that
$\calm(r_0)=4\pi r^3_0 \rho_{0,i}/3.$
%The absence of singularities in the metric functions requires 
%the continuity of $f(r)$. 
As we assume that the homogeneous metric is flat,
we also have $f(r_0)=0$. 
Discontinuities in $f'(r)$ result in 
discontinuities in the derivatives of the metric functions. 
%Alternatively, one may consider the matching of the two metrics at the surface 
%with $r=r_0$ employing junction conditions \cite{israel}. 
%If this surface does not contain a singular energy density, 
%the above constraints must be imposed \cite{biswas2,matching}.

In our modelling we assume that
at the initial time $t_i=0$ the expansion rate $H_i=\dot{R}/R=\dot{R}'/R'$
is given for all $r$ by the standard
expression in homogeneous cosmology:
$H_i^2=\rho_{0,i}/(6M^2)$.
Then, eq. (\ref{tb1}) with $R(0,r)=r$ implies that
\be
f(r)=\frac{\rho_{0,i}}{6M^2}r^2\left(
1-\frac{3\calm(r)}{4\pi r^3 \rho_{0,i}}\right).
\label{fr} \ee
%\be
%^{(3)}R(r,t)=-2 \frac{(fR)'}{R^2 R'}.
%\label{spcurv} \ee
For our choice of $f(r)$, 
%the spatial curvature of the LTB geometry 
%at the initial time is \cite{brouzakis}
%\be
%^{(3)}R(r,0)
%=
%-6 H^2_i
%\left(1- \frac{\calm'}{4\pi  r^2 \rho_{0,i}} \right)
%=-6H^2_i
%\left( 1-\frac{\rho_i(r)}{\rho_{0,i}} \right).
%\label{spcurvin} \ee
overdense regions have positive spatial curvature and tend to contract, 
while underdense ones
negative curvature and expand faster than the average.
This is very similar to the initial condition considered
in the model of spherical collapse.
Even though we work with the particular choice (\ref{fr})
for $f(r)$, we expect that our conclusions are valid for other variations of
the LTB metric as well. These may include an arbitrary function $t_0(r)$ resulting
from the integration of eq. (\ref{tb1}). As this function appears in the combination
$t-t_0(r)$, it becomes irrelevant for large times. Also, the
radial coordinate $r$ is often redefined so that $\rho_i$ is constant. As this is 
only a gauge choice, we do not expect it to affect the physical behaviour. 
The eventual collapse or 
fast expansion of a certain region would be determined by its spatial
curvature, as in our model.  
\\

{\bf Optical equations}:
%Without loss of generality we consider
%geodesic null curves on the plane with $\theta=\pi/2$.
%The first geodesic equation is
%\be
%\frac{dk^0}{d\lx}+\frac{\dot{R}'R'}{1+f}\left(k^1\right)^2+
%\dot{R}R\left( k^3\right)^2=0,
%\label{gtb3}\ee
%with $k^i={dx^i}/{d\lambda}$ and $\lambda$ an affine parameter along the
%null beam trajectory. The second geodesic equation can be replaced by the
%null condition
%\begin{equation}
%-\left(k^0\right)^2+\frac{R'^2}{1+f}\left(k^1\right)^2+R^2
%\left( k^3\right)^2=0,
%\label{gtb4}
%\end{equation}
%while the third one can be integrated to obtain
%$k^3={c_{\phi}}/{R^2}$.
The optical equations \cite{sachs} can be written as \cite{brouzakis}
\begin{eqnarray}
\frac{1}{\sqrt{A}}\frac{d^2\sqrt{A}}{d\lx^2} &=&
-\frac{1}{4M^2}\rho \left( k^0\right)^2  -\sigma^2
\label{exx3} \\
\frac{d \sx}{d \lx} +\frac{2}{\sqrt{A}} \frac{d\sqrt{A}}{d\lx}\, \sx
&=& \frac{\left( k^3 \right)^2 R^2}{4M^2}
\left(\rho -\frac{3 \calm(r)}{4 \pi R^3} \right),
\label{exx33} \end{eqnarray}
where $A$ is the cross section of a light beam, $\lx$ an affine parameter
along the null trajectory and $k^i={dx^i}/{d\lambda}$.
The shear $\sx$ is important when the beam passes near regions in which the
density exceeds the average one by several orders of magnitude.
Within our modelling of large-scale structure, applicable for scales above
${\cal O} (10)\, h^{-1}$ Mpc,
the average density contrast is not sufficiently large for the shear to become
important \cite{brouzakis}. 
%For this reason we neglect it. 

%The FRW metric is a special case of the LTB metric with
%$R(t,r)=a(t)r$, $f(r)=cr^2$, $c=0,\pm 1$ and
%$\rho={c_\rho}/{a^{3}(t)}$.
%The geodesic equation (\ref{gtb3}) has the solution
%$k^0={c_t}/{a(t)}$.
%The solution of eq. (\ref{exx3}) for an outgoing
%beam is
%$A(\lx) = r^2(\lx)\, a^2(t(\lx)) \, \Omega_s.$
We assume that, even for general backgrounds, the light emission near the source is not
affected by the large-scale geometry. By choosing an affine parameter
that is locally $\lx=t$ in the vicinity of the source, we can set
$\left.{d\sqrt{A}}/{d\lx} \right|_{\lx=0}=\sqrt{\Omega_s}.$
The constant $\Omega_s$ can be identified with the solid angle spanned by a
certain beam when the light is emitted by a point-like isotropic source.
This expression, along with
$\left. \sqrt{A} \right|_{\lx=0}=0$,
provide the initial conditions for the solution of eq. (\ref{exx3}).

In order to define the luminosity distance, we consider photons
emitted within a solid angle $\Omega_s$
by an isotropic source with luminosity $L$.
These photons are detected
by an observer for whom the light beam
has a cross-section $A_o$.
The redshift factor is
$1+z={\omega_s}/{\omega_o}={k^0_s}/{k^0_o}$,
because the frequencies measured at the source and at the observation point
are proportional to the values of $k^0$ at these points.
The luminosity distance is 
$D_L=(1+z)\sqrt{A_o/\Omega_s}$, with $A_o$ the beam area measured 
by the observer
for a beam emitted by the source within a solid angle $\Omega_s$. 
The beam area can be calculated by solving
eq. (\ref{exx3}).

It is convenient to switch to dimensionless variables.
We define $\ttb=t H_i$, $\rb=r/r_0$, $\Rb=R/r_0$, where
$H^2_i=\rho_{0,i}/(6M^2)$ is the initial homogeneous
expansion rate and $r_0$ gives the size of the inhomogeneity in
comoving coordinates.
The evolution equation becomes
\be
\frac{\dot{\Rb}^2}{\Rb^2}=\frac{3\calmb(\rb)}{4\pi\Rb^3}+
\frac{\fb(\rb)}{\Rb^2},
\label{eind} \ee
with $\bar{\calm}=\calm/(\rho_{0,i}r^3_0)$
and $\fb=6M^2f/(\rho_{0,i}r_0^2)=f/\Hb^2_i$, $\Hb_i=H_ir_0$.
The dot now denotes a derivative with respect to $\ttb$.
We take the affine parameter $\lx$ to have the dimension of time and
we define the dimensionless variables
$\lb=H_i\lx $,
$\kb^0=k^0$, $\kb^1=k^1/\Hb_i$, $\kb^3=r_0 k^3$.
The geodesic equations maintain their form, with
the various quantities replaced by barred ones, and the combination
$1+f$ replaced by $\Hb_i^{-2}+\fb$. The optical equations
take the form
\begin{eqnarray}
\frac{1}{\sqrt{\Ab}}\frac{d^2\sqrt{\Ab}}{d\lb^2}&=&
-\frac{3}{2}\rhb \left( \kb^0\right)^2 -\sb^2
\label{exx3r}\\
\frac{d \sb}{d \lb} +\frac{2}{\sqrt{\Ab}} \frac{d\sqrt{\Ab}}{d\lb}\, \sb
&=& \frac{3}{2}\left( \kb^3 \right)^2 \Rb^2
\left(\rhb -\frac{3 \bar{\calm}}{4 \pi \Rb^3} \right),
\label{exx33r} \end{eqnarray}
with $\rhb=\rho/\rho_{0,i}$ and $\sb=\sx/H_i$. 
%We omitted the shear, because it gives a negligible contribution
%to our results, as we explained above.
The initial conditions become
$\left. {d\sqrt{\Ab}}/{d\lb} \right|_{\lb=0}=\sqrt{\Omb_s}/\Hb_i
=\sqrt{\Omega_s}$ and 
$\left. \sqrt{\Ab} \right|_{\lx=0}=0$,
with $\Ab=H_i^2A$ and $\Omb=\Hb_i^2 \Omega$.

The effect of the inhomogeneity on the characteristics of the light beam can be
calculated analytically for perturbations with size much smaller than the distance to
the horizon. These have $\Hb_i \ll 1$. In the following we use $\Hb_i$ as a small 
parameter in a perturbative calculation of the luminosity distance and redshift.
For small inhomogeneities, the variation of the Hubble parameter during the crossing by 
the light beam is very small. As a result $\Hb_i$ is almost identical with 
the value $\Hb$ at the time of detection of the beam.
We consider beams with $k^3=0$ that pass through
the center of the spherical inhomogeneity. Beams with $k^3\not= 0$ can also be 
considered along the same lines, even though the calculation is much more involved.
\\

{\bf Travelling time and redshift}:
%The null condition (\ref{gtb4}) can be written as
%\be
%\frac{d\ttb}{d\rb}=\mp \Hb_i\frac{\Rb'}{\sqrt{1+\Hb^2_i \fb}}
%\simeq \mp \Hb_i \Rb' \pm \frac{\Hb^3_i}{2} \Rb'\fb,
%\label{null}\ee
%for incoming and outgoing beams respectively.
%If we keep terms up to ${\cal O}(\Hb^2_i)$ we can neglect the second
%term in the above expression. We can also employ the approximation
%$\Rb'(\ttb,\rb)\simeq \Rb'(\ttb_s,\rb)+\dot{\Rb}'(\ttb_s,\rb)(\ttb-\ttb_s)$,
%as the time it takes for the light to cross the inhomogeneity is much shorter than
%the Hubble time. In fact, $\ttb-\ttb_s={\cal O}(\Hb_i)$.
The travelling time for a beam that propagates across the inhomogeneity has been
calculated in ref. \cite{brouzakis} up to ${\cal O}(\Hb^2_i)$. 
We denote by $\rb_s$ the location of the source and by
$\ttb_s$ the emission time of the beam.
The travelling time is
\begin{eqnarray}
\ttb-\ttb_s=&&\pm\Hb_i \left( \Rb(\ttb_s,\rb_s)-\Rb(\ttb_s,\rb)\right)
+\Hb_i^2\int^{\rb_s}_\rb \Rb'(\ttb_s,\rb) \dot{\Rb}(\ttb_s,\rb) d\rb
\nonumber \\
&&-\Hb_i^2 \left( \Rb(\ttb_s,\rb_s)-\Rb(\ttb_s,\rb)\right) \dot{\Rb}(\ttb_s,\rb)
+{\cal O}(\Hb_i^3)
\label{sol1}
\end{eqnarray}
for incoming and outgoing beams, respectively. The leading term in the
above expression,
of ${\cal O}(\Hb_i)$, is the standard Doppler shift. It is non-zero whenever the
observer has a peculiar velocity relative to a source in the homogeneous region.

We can make a comparison with the propagation of light in a FRW background.
In this case we have $\Rb(\ttb,\rb)=a(\ttb)\rb=\Rb(\ttb,1)\rb$.
%We have expressed the scale factor in terms of the value of the function $\Rb(\ttb,\rb)$
%at the boundary of the inhomogeneous region $\rb=1$.
Let us consider light signals emitted at $\rb_s=1$ and observed at the center
($\rb_o=0$) of the inhomogeneity. The peculiar velocity of such an observer is
zero and the term of ${\cal O}(\Hb_i)$ vanishes.
The difference in propagation time within the
LTB and FRW backgrounds is
\begin{equation}
\ttb_o-\left(\ttb_o\right)_{FRW}=
\Hb_i^2\int_{0}^1 \Rb'(\ttb_s,\rb) \dot{\Rb}(\ttb_s,\rb) d\rb
-\frac{\Hb_i^2}{2} \Rb(\ttb_s,1) \dot{\Rb}(\ttb_s,1)
+{\cal O} (\Hb^3_i).
\label{dt1}
\end{equation}
For signals originating at $\rb_s=0$ and detected at $\rb_o=1$
the time difference has the opposite sign. As
a result, the time difference for signals that cross the
inhomogeneity is of ${\cal O} (\Hb^3_i)$. 

A similar expression can be derived for the redshift of a light beam that
passes through the center of the inhomogeneity.
%The geodesic equation (\ref{gtb3}) can be written as
%\be
%\frac{1}{k^0} \frac{dk^0}{d\rb}=
%-\frac{d\,\ln(1+z)}{d\rb}
%=\pm \Hb_i\frac{\dot{\Rb}'}{1+\Hb^2_i \fb}
%\simeq \pm \Hb_i\dot{\Rb}',
%\label{redshift}\ee
%for incoming and outgoing beams respectively.
%In this way 
One finds \cite{brouzakis2}
\begin{eqnarray}
\ln(1+z)=&&\pm\Hb_i \left( \dot{\Rb}(\ttb_s,\rb_s)-\dot{\Rb}(\ttb_s,\rb)\right)
\nonumber \\
&&+\Hb_i^2\int_{\rb}^{\rb_s} \ddot{\Rb}'(\ttb_s,\rb)
\left(\Rb(\ttb_s,\rb_s)-\Rb(\ttb_s,\rb) \right)d\rb
+{\cal O}(\Hb_i^3)
\label{solred1}
\end{eqnarray}
for incoming and outgoing beams, respectively.

For signals originating at $\rb_s=1$ and detected at $\rb_o=0$
the redshifts obey
\begin{eqnarray}
\ln \left(\frac{1+z}{1+z_{FRW}} \right)=
&&\Hb_i^2\int_0^1 \ddot{\Rb}'(\ttb_s,\rb)
\left(\Rb(\ttb_s,1)-\Rb(\ttb_s,\rb) \right)d\rb
\nonumber \\
&&-\frac{\Hb_i^2}{2} \ddot{\Rb}'(\ttb_s,1)\Rb(\ttb_s,1)
+{\cal O}(\Hb_i^3).
\label{relred} \end{eqnarray}
For signals originating at $\rb_s=0$ and detected at $\rb_o=1$
the r.h.s. of the above equation has the opposite sign. As
a result, the redshift difference for signals that cross the
inhomogeneity is of ${\cal O} (\Hb^3_i)$. 
\\

{\bf Beam area}:
The beam area obeys the second-order differential equation (\ref{exx3r}),
whose solution depends crucially on the initial conditions. In certain
situations, the symmetry of the problem permits an exact solution. 
For example, for signals emitted from some point $\rb_s$ at a
time $\ttb=\ttb_s$ and observed at $\rb_o=0$
we have \cite{partovi}
\begin{equation}
\sqrt{\Ab}= (1+z)\Rb(\ttb_s,\rb_s)\sqrt{\Omb}.
\label{sqrt1} \end{equation}
Similarly, for signals emitted from the center $\rb_s=0$
and observed at $\rb_o$ at a time $\ttb_o$
we have
\begin{equation}
\sqrt{\Ab}= \Rb(\ttb_o,\rb_o)\sqrt{\Omb}.
\label{sqrt2} \end{equation}
However, for a signal that crosses the inhomogeneity we need to integrate
eq. (\ref{exx3r}) from $\rb=0$ to $\rb_o$
with initial conditions determined by the propagation from $\rb_s$ to $\rb=0$.
These include not only $\sqrt{\Ab}$, but $d\sqrt{\Ab}/d\rb$ as well.
An exact analytical solution is not possible in this case, and we have to resort to
perturbation theory in $\Hb_i$. 
We have checked that the expressions 
(\ref{sqrt1}) and (\ref{sqrt2}) 
are reproduced correctly by our results, up to second order in $\Hb_i$.

%Using perturbation theory, 
%it is possible to obtain an analytical estimate 
%of the deviation of the
%luminosity distance from its value in homogeneous cosmology. 
The optical equations (\ref{exx3r}), (\ref{exx33r}) can be written in the form 
\begin{eqnarray}
\frac{d^2\sqrt{\Ab}}{d\rb^2}
+\frac{1}{\left( \kb^1\right)^2}\frac{d\kb^1}{d\lb}
\frac{d\sqrt{\Ab}}{d\rb}
&=&
-\frac{3}{2}\rhb \left( \frac{\kb^0}{\kb^1}\right)^2 \sqrt{\Ab}
-\left(\frac{\sb}{\kb^1} \right)^2 \sqrt{\Ab}
\label{exx3rr} \\
\frac{d}{d \rb} \left( \frac{\sb}{\kb^1}\right)
+\frac{1}{\left( \kb^1\right)^2}\frac{d\kb^1}{d\lb}
\frac{\sb}{\kb^1}
+\frac{2}{\sqrt{\Ab}} \frac{d\sqrt{\Ab}}{d\rb}\, \frac{\sb}{\kb^1}
&=& \frac{3}{2}\left(\frac{\Rb \kb^3}{\kb^1} \right)^2
\left(\rhb -\frac{3 \bar{\calm}}{4 \pi \Rb^3} \right).
\label{exx33rr} \end{eqnarray}
The first term in the r.h.s. of eq. (\ref{exx3rr}) 
is of ${\cal O} \left(\Hb_i^2\right)$ because
$\rhb={\cal O}(1)$ and
$\kb^0/\kb^1=d\bt/d\br=\Hb_i\, dt/dr ={\cal O}\left( \Hb_i\right)$.
The term in the r.h.s. of eq. (\ref{exx33rr}) is also of ${\cal O} \left(\Hb_i^2\right)$
because
$\Rb \kb^3/\kb^1=\Hb_i\,R d\phi/dr ={\cal O}\left( \Hb_i\right)$.
As a result, the second term in the r.h.s. of eq. (\ref{exx3rr}) is 
${\cal O}\left( \Hb^4_i\right)$ and, therefore, negligible. 
The shear plays no role, 
except for cases in which the light passes very close to an extremely dense concentration
of mass. At the length scales that we are considering the energy density is smoothly 
distributed, and the shear can be neglected.  
As the first term in the r.h.s. of eq. (\ref{exx3rr}) 
generates the deviations of the luminosity distance from its value
in a homogeneous background, we expect the overall effect to be of 
${\cal O}\left( \Hb_i^2 \right)$. In the following we confirm this expectation
through an explicit calculation, assuming a simplified form of the energy density.

We consider beam trajectories that start at the boundary of the inhomogeneity, 
pass through its center and exit from the other side. These have $\kb^3=0$.
We express $d\bar{k}_1/d\bar{\lambda}$ in eq. (\ref{exx3rr}) 
using the geodesic equation \cite{brouzakis}, and omit the shear. As the
FRW metric is special case of the LTB one, no change of coordinates 
is necessary. In this way we
obtain
\be \frac{d^2\sbA}{d \brr^2}+\left(\pm \frac{2\sc
\dot{\bR}'}{\sqrt{1+\sc^2\fb}}-
\frac{\bR''}{\bR'}+\frac{\sc^2\fb'}{2(1+\sc^2 \fb)} \right) \frac{d
\sbA}{d \brr}=-\frac{3}{2} \bar{\rho} \frac{R'^2}{1+\sc^2 \fb} \sbA,
\label{eq1} \ee 
where the positive sign in the second term corresponds to ingoing
and the negative sign to outgoing geodesics.

%The initial conditions for an ingoing beam 
%can be taken $\sbA(\rb_s=1)=0$, $d\sbA(\rb_s=1) /d\brr=-1$, without loss of generality.
We use the expansion 
\be
\sbA=\sbAz+\sc \sbAf+\sc^2 \sbAs +{\cal O}(\Hb_i^3),
\ee
and calculate $\sqrt{\Ab^{(i)}}$ in each order of perturbation theory. 
%We point out that our choice of initial configuration, 
%that involves
%a discontinuous energy density, results in the appearance of $\delta$-function singularities 
%in the second derivatives of $\Rb$ with respect to $\rb$. These must be taken into account in
%a consistent calculation. 
The travelling time is given by eq. (\ref{sol1}).
We can set $t_s=0$ so the geodesic inside the inhomogeneity  is
$\bt=-\sc(\brr-1)$
for ingoing, 
and 
$\bt=\sc(\brr+1)$
for outgoing geodesics.
We treat $\bt$ as an ${\cal O}(\sc)$ quantity.
\\

{\bf Central underdensity}:
We identify the initial time in the background evolution
with the time of light emission: $\tb_i=\tb_s=0$. 
This implies that $R'(0,r)=\Rb'(0,\rb)=1$. Also $\dot\Rb'(0,\rb)=1$.  
The initial configuration that we consider has $\bar{\rho}_i(0,\brr)=0$ for
$\brr<\brr_1$ and $\bar{\rho}_i(0,\brr)=1/(1-\brr_1^3)$ for $\brr>\brr_1$.
From (\ref{eind}) we can calculate various derivatives of $\bR$ at
$\bt=0$:
\be \dot{\bR}'(\bt,\brr)=\dot{\bR}'(0,\brr)+\bt
\ddot{\bR}'(0,\brr)+{\cal O}(\sc^2)
=1+\bt
\ddot{\bR}'(0,\brr)+{\cal O}(\sc^2), \ee 
\be \frac{\bR''}{\bR'}(\bt,\brr)=
\frac{\bt^2}{2}\ddot{\bR}''(0,\brr)+{\cal O}(\sc^3). \ee 
For $\brr>\brr_1$ we
have 
\be \ddot{\bR}'(0,\brr)=\frac{r^3+2 \brr_1^3}{2 r^3
\left(\brr_1^3-1\right)}, 
~~~~~~~~~
\ddot{\bR}''(0,\brr)=-\frac{3 \brr_1^3}{r^4
\left(\brr_1^3-1\right)}. \ee 
For $\brr<\brr_1 $ both
$\ddot{\bR}'(0,\brr)$ and $\ddot{\bR}''(0,\brr)$ are zero.
For the initial configuration that we assume, $\ddot{\bR}$ is a continuous function of 
$\rb$. However, $\ddot{\bR}'$ is discontinuous at $\rb=\rb_1$ and $\rb=1$, while 
$\ddot{\bR}''$ has $\delta$-function singularities at the same points. 

The initial conditions for the solution of eq. (\ref{eq1}) for an ingoing beam 
can be taken $\sbA(1)=0$, $d\sbA(1) /d\brr=-1$, without loss of generality.
To zeroth order in $\sc$, eq. (\ref{eq1})
becomes ${d^2 \sbAz}/{d \brr^2}=0$, with solution
$\sbAz(\brr)=-(r-1)$ for ingoing and
$ \sbAz(\brr)=r+1$ for outgoing beams.
To first order in $\sc$, eq. (\ref{eq1})
gives ${d \sbAf}/{d \brr}=-2$, with solution
$\sbAf(\brr)=r^2-2r+1$ for ingoing and 
$\sbAf(\brr)=r^2+2r+1$ for outgoing beams.
These results are the same as for the case of a homogeneous background.

The effect of the inhomogeneity appears in 
second order in $\sc$. We obtain
\bea \frac{d^2\sbAs}{d \brr^2}+\Biggl(\pm 2 \bt(\brr)\ddot{\bR}'(0,\brr)-
\frac{\bt^2}{2}\ddot{\bR}''(0,\brr)&+&\frac{\fb'(\brr)}{2} \Biggr) \frac{d
\sbAz}{d \brr}\pm 2\frac{d
\sbAf}{d \brr}
\nonumber \\
&=&-\frac{3}{2} \bar{\rho}(0,\brr)  \sbAz,
\label{eqqq1} \eea
with the upper sign corresponding to ingoing and the lower one to outgoing
geodesics.
As we have already mentioned, for $\brr<\brr_1$ we have 
$\bar{\rho}_i(0,\brr)= \ddot{\bR}'(0,\brr)=\ddot{\bR}''(0,\brr)=0$.

The above equation can be solved analytically through simple integration, with the 
values at the end of each interval determining the initial conditions for the
next one. The only non-trivial point is that the
$\delta$-function singularities of $\ddot{\bR}''$ at $\rb=\rb_1$ and $\rb=1$ 
induce discontinuities in the values of ${d\sbAs}/{d \brr}$ at these points. These must
be taken into account in a consistent calculation. The discontinuities can be
easily determined through the integration of eq. (\ref{eqqq1}) in an infinitesimal
interval around each of these points. 
The remaining calculation is straightforward.
It must be emphasized that the discontinuous density profiles that we are considering
can be viewed as limiting cases of continuous ones, when the transition regions
become infinitesimally thin. The integration of eq. (\ref{eqqq1}) around the corresponding
values of $r$ picks up the leading contributions arising from the 
transition regions. Including these contributions is necessary in order to reproduce
correctly the exact expressions (\ref{sqrt1}), (\ref{sqrt2}).

For a photon beam that starts from the boundary at $\rb=1$, travels through
the center of an underdensity at $\rb=0$, and exits at the
diametrical point with $\rb=1$, we find 
\be
\sbAs(\br=0)=1-\frac{3}{4}\frac{\br_1+1}{
   \br_1^2+\br_1+1}
%~~~~~
%\sbAs'(\br=0)=-\frac{5}{2}+\frac{3}{2}\frac{1}{
%   \br_1^2+\br_1+1}, 
\ee
and
\be
\sbAs(\br=1)=5-\frac{3}{\brr_1^2+\brr_1+1}.
%~~~~~
%\sbAs'(\br=1)=6-\frac{3}{\brr_1^2+\brr_1+1}.
\ee

Putting everything together, we find that, when the photon exits the inhomogeneity
at $\brr=1$,
\be
\sbA(\brr=1)=2+4\sc+\left(5-\frac{3}{\brr_1^2+\brr_1+1}\right) \sc^2
+{\cal O}(\Hb^3_i).
%\label{final} \\
%\frac{d\sbA}{dr}(\brr=1)&=&1+4\sc+\left(6-\frac{3}{\brr_1^2+\brr_1+1}\right) \sc^2
%+{\cal O}(\Hb^3_i).
\label{finalp} \ee
The expressions for a homogeneous universe are obtained by setting $\rb_1=0$.
The beam area and the luminosity distance are increased by the
presence of the inhomogeneity ($\rb_1\not= 0$). 

We also mention that, if the beam is emitted at
$\br=0$, it exits the inhomogeneity with
$\sbAs(\br=1)={1}/{4}$ and 
$\sbAs'(\br=1)={3}/{4}$, in agreement with eq. (\ref{sqrt2}).
%(both for underdense and
%overdense regions)
\\

{\bf Central overdensity}:
The initial configuration that we consider has $\bar{\rho}_i(0,\brr)=1/\brr_1^3$ for
$\brr<\brr_1$ and $\bar{\rho}_i(0,\brr)=0$ for $\brr>\brr_1$.

For $\br>\br_1$ we
have 
\be \ddot{\bR}'(0,\br)=\frac{1}{\br^3}, 
~~~~~~~~
\ddot{\bR}''(0,\br)=-\frac{3}{\br^4}, \ee 
while for $\br<\br_1$ we
have 
\be 
\ddot{\bR}'(0,\br)=-\frac{1}{2 \br_1^3},
~~~~~~~~
\ddot{\bR}''(0,\br)=0. 
\ee
The expressions for $\sqrt{\Ab^{(0)}}$ and $\sqrt{\Ab^{(1)}}$ are the 
same as in the case of a central underdensity, as they are not affected
by the inhomogeneity. For $\sqrt{\Ab^{(2)}}$
we find
\be 
\sbAs(\br=0)=1-\frac{3}{4}\frac{1}{\br_1}
%~~~~~~~~
%\sbAs'(\br=0)=-\frac{5}{2}+\frac{3}{2}\frac{ \br_1+\frac{1}{6}}{ \br_1^3},
\label{eq2}\ee 
and
\be \sbAs(\br=1)=5-\frac{3}{\br_1^2}.
%~~~~~~~~
%\sbAs'(\br=1^{+})=6-\frac{3}{\br_1^2}. 
\ee 
%when the beam exits the inhomogeneity.

Putting everything together, we find that, when the photon exits the inhomogeneity
at $\brr=1$,
\be
\sbA(\brr=1)=2+4\sc+\left(5-\frac{3}{\brr_1^2}\right) \sc^2
+{\cal O}(\Hb^3_i).
\label{finalo} 
%\\
%\frac{d\sbA}{dr}(\brr=1)&=&1+4\sc+\left(6-\frac{3}{\brr_1^2}\right) \sc^2
%+{\cal O}(\Hb^3_i).
\label{finalpo} \ee
The expressions for a homogeneous universe are obtained by setting $\rb_1=1$.
In this case, the beam area and the luminosity distance are reduced by the
presence of the inhomogeneity ($\rb_1\not= 1$). 
The singularity for $\rb_1\to 0$ is an artifact of the perturbative expansion.
Clearly, the expansion in $\Hb_i$ breaks down when the coefficient of $\Hb_i^2$
diverges. 

The increase of the beam area by a central underdensity with a certain 
$\rb_1$ can always be compensated by the decrease because of an overdensity
with a different value $\rb'_1$. If one requires that $\rb_1$ and $\rb'_1$ be equal, 
the solution is $\rb_1=\rb'_1=2^{-1/3}$. In this case, the central underdensity and 
its surrounding overdense shell, as well as the compensating central overdensity 
and its surrounding underdense shell, all have equal volumes.

If the beam is emitted at
$\br=0$, it exits the inhomogeneity with
$\sbAs(\br=1)={1}/{4}$ and 
$\sbAs'(\br=1)={3}/{4}$, exactly as in the case of a central underdensity.
%(both for underdense and
%overdense regions)
\\

{\bf Flux conservation}:
We have seen that,
when a light beam crosses a certain inhomogeneity,
the deviations of the travelling time $\tb_o$ and redshift $z$ 
from their values in a homogeneous background
are of ${\cal O}(\Hb_i^3)$, 
while the deviation of $\Ab$ is of ${\cal O}(\Hb_i^2)$.
As a result, the effect on the luminosity distance is of
${\cal O}(\Hb_i^2)$. 
%This conclusion has been confirmed through numerical solutions of the
%exact optical equations, without any approximations. Several beam crossings
%were studied for a multitude of values of $\Hb_i$. The 
%coordinate time $\tb_o$ needed for the crossing, the redshift $z$ and the
%beam area ${\Ab}$ were plotted in terms of $\Hb_i$. 
This conclusion holds for any beam going through the inhomogeneity, even if the
crossing is not central. The analytical estimate 
has been verified through the numerical solution
of the optical equations \cite{brouzakis,brouzakis2}. In particular, a central
crossing of a void-like inhomogeneity (with a central underdensity) 
results in the increase of the 
luminosity distance by an amount of ${\cal O}(\Hb_i^2)$ \cite{brouzakis,brouzakis2}. 
This result is in agreement with the analysis of ref. \cite{marra}, in which
a sequence of central crossings is assumed during the propagation of
light from source to observer. On the other hand, if the inhomogeneity is
crossed through the overdense region near its surface a decrease of
the luminosity distance by an amount of ${\cal O}(\Hb_i^2)$ takes place 
\cite{brouzakis,brouzakis2}.

The conclusion that the redshift is affected by an amount of
${\cal O}(\Hb_i^3)$ has a very important implication.
If the redshift is not altered significantly by the propagation in the inhomogeneous
background, the conservation of the total flux requires that the average luminosity
distance be the same as in the homogeneous case. 
The energy flux may be redistributed in various directions but the total
flux must be the same as in the homogeneous case \cite{weinberg,rose}. 
The maximal deviation from exact flux conservation is 
determined by the effect of the inhomogeneity on the redshift, 
which is of ${\cal O}(\Hb_i^3)$.
As a result, even though the effect on the luminosity distance for a single 
crossing is of ${\cal O}(\Hb_i^2)$, the 
{\it maximal average} effect for beams originating in
the same source and crossing the inhomogeneity at various angles is
of ${\cal O}(\Hb_i^3)$. 
In the case of an underdensity, the increase of the luminosity distance 
for central beam crossings is compensated by a reduction for beams that
travel mainly through the peripheral overdense shell. The opposite 
happens in the case of a central overdensity.

The above conclusion has been verified numerically in ref. 
\cite{brouzakis2}, both for central underdensities and overdensities. 
An equivalent conclusion is that the maximal statistical effect for
light signals received from randomly distributed sources in the sky should
be of ${\cal O}(\Hb_i^3)$, similarly to the effect on the redshift.
The statistical analysis of ref. \cite{brouzakis2} confirms this
expectation.
\\

{\bf Conclusions}:
The effect of spherical inhomogeneities on light emitted by a distance source 
depends on $\Hb=r_0 H$. 
For an observer located at the center of a spherical inhomogeneity, the
deviations of travelling time, redshift, 
beam area and luminosity distance from their values
in a homogeneous background are of ${\cal O}(\Hb^2)$. The luminosity 
distance is increased by the presence of a central underdensity, while it is
reduced by a central overdensity. 
The increase in the luminosity distance if the observer is 
located near the center of a large void
can by employed for the explanation of the supernova data \cite{central}.
An increase of ${\cal O}(10\%)$, as required by the data, would imply the 
existence of a void with size of ${\cal O}(10^3)\, h^{-1}$ Mpc. 
Numerical factors can reduce the required size, depending on the details of
the particular cosmological model employed \cite{central}. However, a typical void 
with size of ${\cal O}(10)\, h^{-1}$ Mpc leads to a negligible 
increase of the luminosity distance. 

If the observer is located at a random position within the
homogeneous region, the beam can cross several 
inhomogeneities before its detection. Each crossing produces an 
effect of 
${\cal O}(\Hb^3)$ for the travelling time and the redshift.
For the beam area and the luminosity distance the effect 
is of ${\cal O}(\Hb^2)$.  
However, flux conservation implies
that positive and negative contributions to the beam area 
cancel during multiple crossings. 
The size of the {\it maximal average} effect of each crossing on
the beam area and luminosity distance is set by the effect on
the redshift, which is of ${\cal O}(\Hb^3)$ \cite{weinberg,rose,brouzakis2}. 
Photons with redshift $\sim 1$ pass through $\sim (1/H)/r_0=\Hb^{-1}$ inhomogeneities
before arrival, assuming that these are tightly packed. 
As a result, the expectation is that the maximal final effect for a random
position of the observer is of 
${\cal O}(\Hb^2)$ for all quantities. 
This conclusion is
supported by the numerical analysis \cite{brouzakis,brouzakis2}.

We mention at this point that, even for a random position of the observer, 
there is a bias in the residual effect 
on the luminosity distance for a limited sample of sources. The bias is 
towards increased values 
if the Universe is dominated by void-like configurations.
We did not discuss this point in this letter, as we assumed that
the data sample is large. 
A detailed study can be found in ref. \cite{brouzakis2}, to which
we refer the reader for the details.

We conclude that the presence of spherical inhomogeneities 
does not influence sufficiently 
the propagation of light in order to provide an explanation for the supernova
data, unless their size becomes comparable to the horizon distance. 
It is possible, however, that relaxing the assumption 
of spherical symmetry for the inhomogeneities
may increase the influence of the local geometry 
on the beam characteristics and provide an effect at a lower order in $\Hb$. 
The crucial question is whether the influence of the inhomogeneities 
on the redshift can become
larger than the effect of ${\cal O}(\Hb^3)$ predicted by our model and 
the Rees-Sciama estimate \cite{rees}. The modelling
of the Universe as an ensemble of inhomogeneities, glued together by
a homogeneous region (the "Swiss-cheese" model), may be too constraining.
Photons that cross an inhomogeneity enter an
evolving newtonian potential from a homogeneous region, to which they 
subsequently return. Within this modelling, the residual effect 
cannot be much larger than of ${\cal O}(\Hb^3)$. The elimination of the
intermediate homogeneous region may be necessary in order to produce 
a larger effect.
This possibility poses formidable technical difficulties, but 
merits further investigation.

%\vspace {0.5cm}
{\it Acknowledgments}:
%\noindent
This work was supported by the research programs
``Kapodistrias'' of the University of Athens
and ``Pythagoras II'' (grant 70-03-7992)
of the Greek Ministry of National Education, partially funded by the
European Union.

\end{document}